\documentclass[superscriptaddress,twocolumn,showpacs,showkeys,
amssymb,amsmath,nobibnotes,aps,prd,
nofootinbib]{revtex4-1}
\pdfoutput=1
\usepackage{graphicx,subfigure,bm,color,psfrag,hyperref}
\usepackage{amsfonts}
\usepackage{lipsum}
\usepackage{mathtools}
\usepackage{verbatim}
\usepackage[normalem]{ulem}
\usepackage[dvipsnames]{xcolor}
\definecolor{cardinal}{rgb}{0.77, 0.12, 0.23}
\definecolor{lightcardinal}{rgb}{0.97, 0.42, 0.53}
\definecolor{lightblue}{rgb}{0.66, 0.84, 0.96}
\hypersetup{colorlinks,linkcolor={blue},citecolor={red},urlcolor={cardinal}}

\begin{document}

\title{Reheating constraints and the $H_0$ tension  in Quintessential Inflation}

\author{Jaume de Haro}
\email{E-mail: jaime.haro@upc.edu}
\affiliation{Departament de Matem\`atiques, Universitat Polit\`ecnica de Catalunya, Diagonal 647, 08028 Barcelona, Spain}

\author{Supriya Pan}
\email{E-mail: supriya.maths@presiuniv.ac.in}
\affiliation{Department of Mathematics, Presidency University, 86/1 College Street,  Kolkata 700073, India}
\affiliation{Institute of Systems Science, Durban University of Technology, PO Box 1334, Durban 4000, Republic of South Africa}

\thispagestyle{empty}

\begin{abstract}

In this work, we focus on two important aspects of modern cosmology: reheating and Hubble constant tension within the framework of a unified model, namely,  quintessential inflation connecting the early inflationary era and late-time cosmic acceleration. In the context of reheating,  we use instant preheating and gravitational reheating, two viable reheating mechanisms when the evolution of the universe is not affected by an oscillating regime. After obtaining the reheating temperature, we analyze the number of $e$-folds and establish its relationship with the reheating temperature. This allows us to connect, for different quintessential inflation models, the reheating temperature with the spectral index of scalar perturbations, thereby enabling us to constrain its values. In the second part of this article, we explore various alternatives to address the $H_0$ tension, a discrepancy which indicates a possible revision of the $\Lambda$CDM model. Initially, we establish that quintessential inflation alone cannot mitigate the Hubble tension by solely deviating from the concordance model at low redshifts. The introduction of a phantom fluid, capable of increasing the Hubble rate at the present time, becomes a crucial element in alleviating the Hubble tension, resulting in a deviation from the $\Lambda$CDM model only at low redshifts. On a different note, by utilizing quintessential inflation as a source of early dark energy, thereby diminishing the physical size of the sound horizon close to the baryon–photon decoupling redshift, we observe a reduction in the Hubble tension. This alternative avenue, which has the same effect of  a cosmological constant changing its scale close to the recombination, 
sheds light on the nuanced interplay between the quintessential inflation and the Hubble tension, offering a distinct perspective on addressing this cosmological challenge.

\end{abstract}

\vspace{0.5cm}


\maketitle



\section{Introduction}

Inflation and late-time cosmic acceleration are two significant eras of modern cosmology that created enormous interest and debates in the scientific community. Inflation---a rapid accelerating phase of the universe \cite{Guth:1980zm,Linde:1981mu}—is a theory for the early universe that was proposed to solve a number of limitations of the hot big bang cosmology. Usually a scalar field with slowly rolling potential can drive this accelerating phase quite smoothly,  and this motivated the early universe community to investigate various scalar field models with similar features~\cite{Barrow:1981pa,Lucchin:1984yf,Burd:1988ss,Barrow:1990nv,Freese:1990rb,Barrow:1993hn,Barrow:1994nt,Parsons:1995ew,Barrow:1995xb,Boubekeur:2005zm,Martin:2006rs,Barrow:2006dh,Barrow:2007zr,Sebastiani:2013eqa,Kehagias:2013mya,Freese:2014nla,Barrow:2016qkh,Ooba:2017ukj,Giare:2019snj,Anber:2020qzb,Aoki:2021aph,Forconi:2021que,Schimmrigk:2021tlv,Bahr-Kalus:2022prj,He:2024wqv,Lorenzoni:2024krn} (also see \cite{Martin:2013tda,Odintsov:2023weg,Ellis:2023wic,Giare:2024akf}). 
According to the cosmic microwave background observations from various surveys \cite{Fixsen:1993rd,Bennett:1996ce,WMAP:2003syu,WMAP:2012fli,Planck:2013jfk,Planck:2015sxf,Planck:2018jri,Planck:2019nip,ACT:2020frw,SPT-3G:2014dbx,SPT-3G:2021eoc,SPT-3G:2022hvq}, this proposal is one of the leading theories for describing the early universe. 
On the other hand, late-time cosmic acceleration was discovered {at the end of the nineties of the past century through}
 the observations from Type Ia supernovae \cite{SupernovaSearchTeam:1998fmf,SupernovaCosmologyProject:1998vns}.  The discovery of the accelerated expansion of the universe gave a massive jerk on the understanding of the dynamical evolution of the universe and introduced the concept of dark energy or gravity modifications.  One of the possible DE candidates is the quintessence \cite{Caldwell:1997ii,Carroll:1998zi,Tsujikawa:2013fta}, a canonical scalar field that was proposed as a possible alternative to the cosmological constant. One can realize  that both the inflation and late-time cosmic acceleration can be described in terms of the scalar field models, but indeed both eras have different energy scales. 
 {Although the origin of these kinds of scalar fields follows a phenomenological route (see \cite{Chimento:2007da,Gao:2009me,Luongo:2023aaq} to find several attempts to physically explain their appearance)}, this motivated us to introduce a new and appealing cosmological theory, known as quintessential inflation.

Quintessential inflation, {a concept elaborated in \cite{Peebles:1998qn}}, is a theoretical framework that unifies the early and late-time accelerated expansions of the universe, proposing a single scalar field responsible for both inflation and dark energy. This approach addresses key cosmological puzzles by merging two distinct eras of cosmic acceleration into one coherent model. {At early times, the inflation field drives rapid inflation, which solves the horizon, flatness, and other  problems} while generating primordial density perturbations. Post-inflation,  that is, after reheating and the radiation and matter domination phases, the field transits to a slow-roll regime, acting as quintessence, which explains the observed late-time acceleration. The evolution of this scalar field is governed by a potential that allows for a transition between these phases, avoiding the need for separate inflation and dark energy mechanisms. Quintessential inflation models are evaluated against observational data, such as cosmic microwave background (CMB) measurements and large-scale structure surveys, to constrain the potential forms and field dynamics, ensuring consistency with the standard cosmological model. We refer to an incomplete list of works in this direction~\cite{Peloso:1999dm,Giovannini:1999qj,Kaganovich:2000fc,Dimopoulos:2000md,Yahiro:2001uh,Dimopoulos:2001ix,Campos:2002yk,Nunes:2002wz,Giovannini:2003jw,Tashiro:2003qp,Sami:2004xk,Rosenfeld:2005mt,Zhai:2005ub,Cardenas:2006py,Membiela:2006rj,Neupane:2007mu,Bento:2009zz,Hossain:2014coa,Geng:2015fla,deHaro:2016hpl,Guendelman:2016kwj, deHaro:2016hsh,deHaro:2016ftq,Geng:2017mic,AresteSalo:2017lkv,Haro:2015ljc,Agarwal:2017wxo,Bettoni:2018pbl,Dimopoulos:2019ogl,Haro:2019peq,Verner:2020gfa,Benisty:2020xqm,Dimopoulos:2020pas,AresteSalo:2021lmp,AresteSalo:2021wgb,Salo:2021vdv,deHaro:2021swo,Karciauskas:2021fdu,Bettoni:2021qfs,Fujikura:2022udt,Dimopoulos:2022rdp,Inagaki:2023mxv,Alho:2023pkl,Giare:2024sdl}.

After the inflation, a reheating mechanism  becomes essential \cite{Kofman:1994rk,Kofman:1997yn,Greene:1997fu}, because following the reheating of the universe, it transits from its cold inflationary state to a hot phase required to match with the standard Big Bang cosmology. Without reheating, we would have a universe without any matter, and this goes against the observational evidence that we have traced out so far.  
In our work, in the context of quintessential inflation, we review two mechanisms to reheat the universe after inflation, namely, instant \mbox{preheating \cite{Felder:1998vq}} and gravitational reheating via particle production~\cite{Parker:1968mv,Zeldovich:1971mw,Grib:1980aih}. And dealing with { super-symmetric gravity theories
such as}
$\alpha$-attractors in quintessential inflation~\cite{Haro:2011zza}, we  relate the reheating temperature  with the
observable parameters of the power spectrum, namely the spectral index and the ratio of tensor to scalar perturbations, showing how the reheating temperature, whose value ranges between $1$ MeV and $10^9$ GeV in order to guarantee the success of Big Bang Nucleosynthesis (BBN) and overpass the gravitino problem \cite{Khlopov:1984pf, Ellis:1984eq},   constraint these parameters.

Now, during the second phase of quintessential inflation, that means in the era of late-time cosmic acceleration, observational data from various astronomical missions is pointing towards a discrepancy in several key cosmological parameters. One of the intriguing cosmic conundrums arises from a notable disparity between two independent measurements of the present value of the Hubble rate, denoted as $H_0$, which represents the {current} rate of cosmic expansion. More precisely, observations from the cosmic microwave background (CMB) within the $\Lambda$CDM model, utilizing the precise measurements from missions such as the Planck satellite, yield a Hubble constant value $H_0=67.4\pm0.5$ km~s$^{-1}$Mpc$^{-1}$~\cite{Planck:2018vyg}. However, the late-time measurements of $H_0$ using the cosmic distance ladder, such as the SH0ES (Supernovae and $H_0$ for the Equation of State (EoS) of dark energy) team, where the Type Ia supernovae are calibrated with Cepheids, lead to $H_0=73.04\pm1.04$~km~s$^{-1}$~Mpc$^{-1}$~\cite{Riess:2021jrx}. These two measurements exhibit a significant tension at the level of $\sim 5\sigma$, posing a challenge to the standard cosmological paradigm and consequently hints at potential shortcomings or missing elements in the $\Lambda$CDM model. {Although there is no doubt that $\Lambda$CDM cosmology has been quite successful in explaining a large number of astronomical surveys, this significant tension in $H_0$ argues that most probably $\Lambda$CDM is an approximate version of a more realistic theory that is under the microscope. }
This tension propels cosmologists to explore alternative models, extensions, or modifications that could provide a more accurate and coherent description of the universe's expansion history. Various avenues have been explored in the literature, including the introduction of new physics, modifications to the nature of dark energy, or reconsideration of fundamental cosmological assumptions~\cite{DiValentino:2021izs,Perivolaropoulos:2021jda,Abdalla:2022yfr}.  {However, based on the existing literature, the final answer is yet to be revealed. }

In the present article, within the framework of quintessential inflation, we delve into two alternative modifications of the concordance model. 
Firstly, we explore the deviation of the concordance model at low redshifts. In this scenario, quintessential inflation alone, without the presence of other components, proves insufficient to reconcile the Hubble tension. However, when combined with quintessential inflation, the introduction of a phantom fluid, altering the dynamics specifically at redshifts $z\lesssim 2$, emerges as a solution. This modification effectively increases the value of the Hubble rate at the present epoch, offering a means to alleviate the tension observed in the Hubble constant measurements.  In the second case, we investigate the injection of dark energy, commonly referred to as early dark energy (EDE). The conceptual foundation involves presenting a potential that, akin to standard quintessential inflation, undergoes a phase transition during the early universe from inflation to kination. During this phase, the potential is nearly flat with a very small scale, mimicking characteristics of a cosmological constant. However, to imbue quintessential inflation with the role of a source of EDE, the potential necessitates another phase transition close to recombination. This secondary transition aims to mimic yet another lower cosmological constant, aligning with the present value of the Hubble rate and offering an alternative approach to mitigate the Hubble tension. In essence, our exploration of these alternative modifications within the quintessential inflation framework unfolds as an endeavor to enhance the concordance model and offer nuanced solutions to the intriguing cosmic puzzle presented by the Hubble tension.

This paper is organized as follows: In Section~\ref{sec-quintessential inflation}, we discuss various quintessential inflation models, starting with the original Peebles–Vilenkin model \cite{Peebles:1998qn} and its companions. In Section~\ref{sec-instant-preheating}, we discuss instant preheating, and in Section~\ref{sec-gravitational-reheating}, we introduce gravitational reheating via particle production. Section~\ref{sec-H0} discusses the Hubble constant  tension in the context of quintessential inflation. Finally, in Section~\ref{sec-summary}, we close the present article, describing the key findings in short.

\section{Quintessential Inflation: The Set-up}
\label{sec-quintessential inflation}

{
Throughout this article, we use the spatially flat Friedmann–Lemaître–Robertson–Walker (FLRW) metric, with $a(t)$ representing the universe's scale factor. The core idea behind the theory of quintessential inflation is as follows: the inflation field is responsible for both the early and late-time acceleration of the universe. After inflation ends, there is a sudden phase transition leading to a period of kination (where the field's energy is purely kinetic, characterized by an EoS $w_{\rm eff}=1$). This breaks the adiabatic evolution, enabling the creation of superheavy particles. The energy density of these particles (scaling as $\rho \sim a^{-3}$) eventually surpasses that of the inflation field (which scales as $\rho_{\varphi} \sim a^{-6}$) once they decay into lighter particles. This process transitions the universe into the radiation-dominated phase of the hot Big Bang. As the universe cools, particles become non-relativistic, leading to matter dominance. Eventually, in the present era, inflation’s energy density rises again in the form of dark energy, known as quintessence, driving the current cosmic acceleration.}

\subsection{The Peebles–Vilenkin Quintessential Inflation}

The first model of {quintessential inflation}, introduced by Peebles and Vilenkin in   \cite{Peebles:1998qn}, contains the following potential:
\begin{eqnarray}\label{PV}
V(\varphi)=\left\{\begin{array}{ccc}
\lambda \left(\varphi^4+  M^4 \right)& \mbox{for} & \varphi\leq 0\\
\lambda\frac{M^8}{\varphi^{4}+M^4} &\mbox{for} & \varphi\geq 0,\end{array}
\right.
\end{eqnarray}
where $\lambda$  is a dimensionless parameter {and $M$ is a very small mass compared with the reduced Planck's mass $M_{pl}$ (i.e., $M\ll M_{pl}$).} Note that the abrupt phase transition occurs at $\varphi=0$, 
where the fourth derivative of $V$ is discontinuous.

{
In this simplified model, the first part of the potential, the quartic term, drives inflation, while the inverse power law potential, forming the quintessence tail, is responsible for the current cosmic acceleration. As discussed in \cite{Peebles:1998qn}, the parameter $\lambda \cong 9\times 10^{-11}$ is determined from the scalar perturbation power spectrum. The value of $M\sim 200$ TeV is derived from observational data, specifically from $\Omega_{\varphi,0}\equiv \frac{\rho_{\varphi,0}}{3H_0^2M_{pl}^2}\cong 0.7$.
}

{Considering that the inflationary component of the potential is quartic, a straightforward calculation reveals the relationship between the number of final $e$-folds (from horizon crossing to the end of inflation) and the spectral index:}
\begin{eqnarray}\label{N-pv}
    N=\frac{6}{1-n_s}-2.
    \end{eqnarray}

According to Planck 2018 data \cite{Planck:2018jri}, the spectral index is measured to be {$n_s=0.9649\pm 0.0042$}. 
{From (\ref{N-pv}) we can see that,  
at the 2$\sigma$ confidence level,  the number of $e$-folds is too high}. It is bound by
$136\leq N\leq 223$,  which shows that the Peebles–Vilenkin model is incompatible with the observational data, and thus it has to be ruled out.

Another way to disregard the model is considering 
the {tensor/scalar ratio}, which   is given by  {$r=\frac{16}{3}(1-n_s)$}. {Therefore, 
at a $2\sigma$ confidence level, the constraint is {$0.1424\leq r\leq 0.232$}, which is inconsistent with the observational limit of $r\leq 0.1$ as reported by \cite{Planck:2018jri}.}

\begin{figure}
\includegraphics[width=0.4\textwidth]{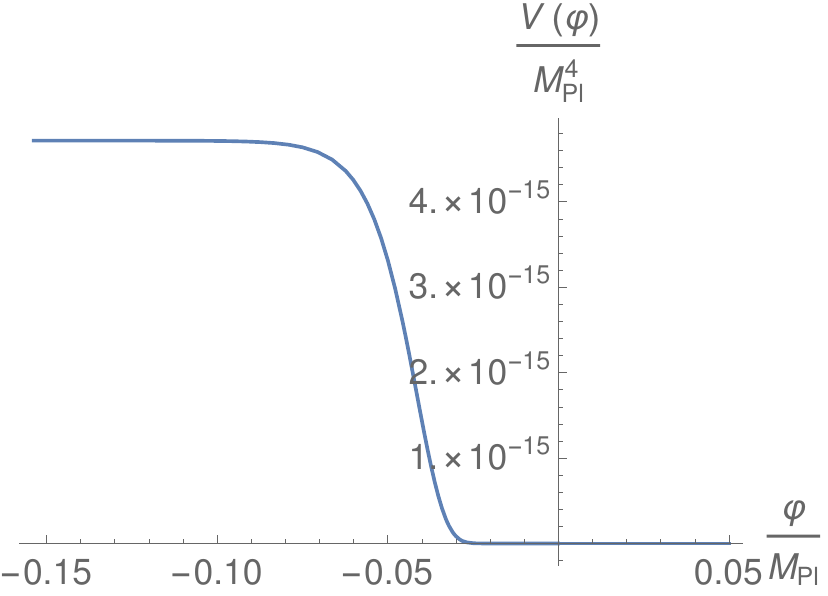}
\caption{The shape of the potential in Lorentzian quintessential inflation. 
 Inflation ends when $\varphi_{\rm END}\cong -0.078M_{pl}$, and kination starts when $H\sim H_{\rm kin}\cong 4\times 10^{-8} M_{\rm pl}$ with $\varphi_{\rm kin}\cong-0.03 M_{\rm pl}$. The figure has been taken from \cite{Salo:2021vdv}.  }
\label{fig:potentialLquintessential inflation}
\end{figure}

\subsubsection{Improved Version: Starobinsky Inflation $+$ Inverse Power Law}

The problem in the Peebles–Vilenkin model is the quadratic potential responsible for inflation. Then, replacing it by a plateau-type potential, for example, the Starobinsky one (see, for instance, \cite{Ivanov:2021chn},  one can obtain spectral values compatible with the observational data. Considering the potential
\begin{eqnarray}\label{PVimproved}
V(\varphi)=\left\{\begin{array}{ccc}
\lambda M_{pl}^4\left(1-e^{\sqrt{\frac{2}{3}}\frac{\varphi}{M_{pl}}}\right)^2 + \lambda M^4 & \mbox{for} & \varphi\leq 0\\
\lambda\frac{M^8}{\varphi^{4}+M^4} &\mbox{for} & \varphi\geq 0,\end{array}
\right.
\end{eqnarray}
one has the relation between the spectral index and the tensor/scalar ratio 
$$r=3(1-n_s)^2 \Longrightarrow  r<5\times 10^{-3}, $$
which matches perfectly with Planck's data.

\subsection{Lorentzian Quintessential Inflation}

Based on the  Lorentzian (Cauchy for Mathematicians) distribution,  one considers the following  {\it ansatz}
 \cite{Benisty:2020qta}:
\begin{eqnarray}\label{ansatz} 
\epsilon(N)=\frac{\xi}{\pi}\frac{\Gamma/2}{N^2+\Gamma^2/4},
\end{eqnarray}
where $\epsilon$ {denotes}  the  main {slow-roll parameter},  $N$ {is once again}  the number of $e$-folds, $\xi$ is the amplitude of the {Cauchy} distribution, and $\Gamma$ is its width. 
{We} obtain the following potential \cite{AresteSalo:2021lmp}:
\begin{eqnarray}\label{Lquintessential inflation}
V(\varphi)=\lambda M_{pl}^4\exp\left[-\frac{2\gamma}{\pi}\arctan\left(\sinh\left(\gamma\varphi/M_{pl} \right)  \right)\right],
\end{eqnarray}
 where $\lambda$ is a dimensionless parameter, and the parameter $\gamma$ is defined by
$\gamma\equiv \sqrt{\frac{\pi}{\Gamma \xi}}.$  In Figure \ref{fig:potentialLquintessential inflation}, the evolution of this potential is depicted. 

The model is governed by two parameters, and in order to align with current observational data, one must set {$\lambda\sim 10^{-69}$} and {$\gamma\cong 122$}. This choice results in a successful inflationary phase and, at late times, leads to eternal cosmic acceleration with an effective EoS parameter equal to $-1$.

\begin{figure}
\includegraphics[width=0.4\textwidth]{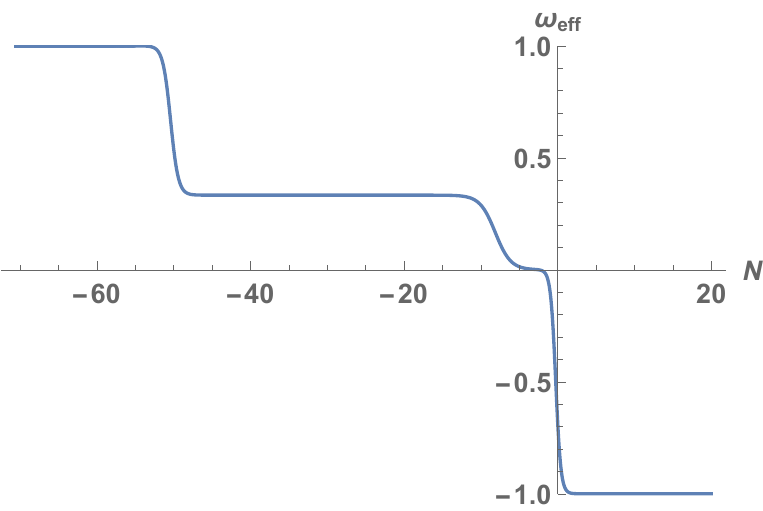}
\caption{Evolution of the effective EoS parameter. The figure has been taken from \cite{AresteSalo:2021wgb}.   }
\label{fig:weff}
\end{figure}
\subsection{$\alpha$-Attractors in Quintessential Inflation}

The Lagrangian, depicting the evolution of a field under the action of an exponential potential,  provided by {super-symmetric} gravity theories, is \cite{Dimopoulos:2017zvq}
\begin{eqnarray}\label{lagrangian}
\mathcal{L}=\frac{1}{2}\frac{\dot{\phi}^2}{(1-\frac{\phi^2}{6\alpha}  )^2}M_{pl}^2-\lambda M_{pl}^4 e^{-\kappa \phi},
\end{eqnarray}
where $\phi$ is a dimensionless scalar field, and $\kappa$ and $\lambda$ are positive dimensionless constants.
In order that the kinetic term has the canonical  form, i.e., 
$\frac{1}{2}{\dot{\varphi}^2}$, one can redefine the scalar field as follows:
\begin{eqnarray}\label{canonical}
\phi= \sqrt{6\alpha}\tanh\left(\frac{\varphi}{\sqrt{6\alpha}M_{pl}}  \right)
\end{eqnarray}
obtaining the following potential {\cite{Dimopoulos:2017zvq,AresteSalo:2021wgb}}: \begin{eqnarray}\label{alpha}
V(\varphi)=\lambda M_{pl}^4e^{-n\tanh\left(\frac{\varphi}{\sqrt{6\alpha}M_{pl}} \right)}, \qquad  n\equiv \kappa\sqrt{6\alpha}.
\end{eqnarray}

In Figure \ref{fig:weff}, we show the evolution of  the effective EoS parameter 
$w_{\rm eff}=-1-\frac{2\dot{H}}{3H^2}$ in terms of  
$N=\ln\left(a/a_0\right)=-\ln (1+z)$ ($a_0$ denotes the present value of the scale factor). {The numerical integration begins at the start of kination, and it is observed at the present time, {$w_{\rm eff}<-1/3$}, indicating that the universe is accelerating. As time progresses, our universe transits to a de Sitter phase, where the equation of state parameter approaches to {$w_{\rm eff}=-1$}.}

\section{Instant Preheating}
\label{sec-instant-preheating}

The idea of instant preheating was presented in~\cite{Felder:1998vq} as a mechanism to reheat the universe in standard inflation. Later, it has been proved that it is very useful in quintessential inflation
\cite{Felder:1999pv}.  Here, we will describe the basic ideas of this reheating mechanism:

{First of all, we consider a quantum field, namely  $\phi$,  responsible for particle production
whose Lagrangian density is given by}
\begin{eqnarray}\label{Lagrangian-density}
{\mathcal L}=\frac{1}{2}\sqrt{|g|}(g^{\mu\nu}\partial_{\mu}\phi
    \partial_{\nu}\phi-\tilde{g}^2(\varphi-\varphi_{\rm kin})^2    \phi^2 \nonumber\\-\xi R\phi^2 +{h\bar{\psi}\psi\phi}),\end{eqnarray}
where  $R$ {denotes}  the {scalar curvature}, $\varphi_{\rm kin}$ is the value of the inflation at the onset of kination, $\tilde g$ is the {dimensionless coupling constant} between the inflation  and the quantum field,
{and $h\bar{\psi}\psi\phi $ is the  usual Yukawa   interaction between the quantum field $\phi$ and fermions $\psi$
\cite{Felder:1999pv}.}
Considering 
{conformally coupled particles}, i.e., {when} $\xi=1/6$, the frequency of the  {$k$-modes is}
\begin{eqnarray}\label{instant0}
\omega_k^2(\eta)=k^2+m_{\rm eff}^2(\eta)a^2(\eta),
\end{eqnarray}
where $m_{\rm eff}(\eta)=\tilde{g}
(\varphi(\eta)-\varphi_{\rm kin})$ is  the {effective mass} of the produced particles, which grows during kination. 
{
In order to calculate analytically the {Bogoliubov coefficients} (the main ingredient to obtain the reheating temperature), 
we perform}
the {linear  approximation}
$\varphi(\eta)-\varphi_{\rm kin}\cong \varphi_{\rm kin}'(\eta-\eta_{\rm kin})$,  and {we assume} that  the  universe is {static} with $a(\eta)=a_{\rm kin}$. Then, the frequency becomes 
\begin{eqnarray}
\omega_k^2(\eta)=k^2+
\tilde{g}^2 a_{\rm kin}^2(\varphi'_{\rm kin})^2(\eta-\eta_{\rm kin})^2.\end{eqnarray}
Thus, using the complex WKB approximation, the analytic value of the $\beta$-Bogoliubov coefficients is given by \cite{Felder:1999pv}
\begin{align}\label{instant}
    |\beta_k|^2\cong \exp \left({-\frac{\pi k^2}{\tilde{g}a_{\rm kin}\varphi'_{\rm kin}}}\right)
    = \exp\left({-\frac{\pi k^2}{\sqrt{6}\tilde{g}a_{\rm kin}^2H_{\rm kin}M_{pl}}}\right).
     \end{align}
Thus, the number density of produced particles at the onset of kination is
\begin{align}
   \langle \hat{N}_{\rm kin}\rangle=\frac{1}{2\pi^2a_{\rm kin}^3}\int_0^{\infty}
   k^2|\beta_k|^2 dk
   = \frac{1}{8\pi^3}(\tilde{g} \sqrt{2\rho_{\rm B,kin}})^{3/2},
    \end{align}
where $\rho_{\rm B, kin}=3H_{\rm kin}^2M_{\rm pl}^2$ is the {background energy density} at the beginning of kination.
Since the effective mass of the produced particles grows, they have to decay, {with a decay rate 
${\Gamma}\equiv \Gamma(\phi\rightarrow \psi\psi)\sim \frac{h^2\tilde{g}M_{\rm pl}}{8\pi}$ (see \cite{Felder:1999pv} for details)}  in lighter ones,
which, after thermalization,  will 
 reheat the universe. After some calculations, the reheating temperature is given by
 \cite{deHaro:2023xcc}
     \begin{eqnarray}
    T_{\rm reh}\cong 3\times 10^{-3}\tilde{g}^{15/8}\left(
\frac{M_{\rm pl} }{\bar\Gamma}
    \right)^{1/4}\ln^{3/4}\left( \frac{M_{\rm pl}}{\bar\Gamma}\right)M_{\rm pl},
\end{eqnarray}     
where we have introduced the notation $\bar\Gamma\equiv 10^7 \Gamma$, $\Gamma$  being the {decay rate}, and we have  taken {$H_{\rm kin}\sim 10^{-7} M_{\rm pl}$}.  To avoid problems such as the {gravitino} problem or the {overproduction of gravitational waves}, 
a successful reheating of the universe ($T_{\rm reh}\leq 10^9$ GeV) is achieved through the instant preheating mechanism when the value of the coupling constant satisfies {$10^{-6}\leq \tilde{g}\leq 3\times 10^{-5}$} and the decay rate is 
{$ 2\times 10^{33}\tilde{g}^{15/2}
    \leq  \frac{\bar\Gamma}{M_{\rm pl}} <1/3 $}.
For example, taking ${\tilde{g}=6\times 10^{-6}}$, we obtain
\begin{eqnarray}\label{reheating-decay}
T_{\rm reh}\cong 5\times 10^{-13}\left(
\frac{M_{\rm pl} }{\bar\Gamma}
    \right)^{1/4}\ln^{3/4}\left( \frac{M_{\rm pl}}{\bar\Gamma}\right)M_{\rm pl},  
\end{eqnarray}
    with $10^{-6}\leq \frac{\bar\Gamma}{M_{\rm pl}} <1/3$ $ \Longrightarrow$      $10^{-13}
    \leq  \frac{h^2\tilde{g}}{8\pi} <\frac{1}{3}\times 10^{-7}
     \Longrightarrow     \frac{4\pi}{3}\times 10^{-7}
    \leq  h^2 <\frac{4\pi}{9}\times 10^{-1}$,
    which {results in the following upper and lower bounds for the reheating temperatures:}
 \begin{eqnarray}\label{max-min}
T_{\rm reh}^{\rm max}\cong  10^{9} \mbox{ GeV} \qquad \mbox{and} \qquad
T_{\rm reh}^{\rm min}\cong 10^{6} \mbox{ GeV}.
\end{eqnarray}

\section{Gravitational Reheating}
\label{sec-gravitational-reheating}

Gravitational reheating \cite{Parker:1968mv, Zeldovich:1971mw,Grib:1980aih,Haro:2011zza,deHaro:2022ukj} is a compelling mechanism by which the interaction of a quantum field coupled with gravity 
generates heavy massive particles that decay
 into standard model (SM) particles and other forms of matter and radiation, thereby reheating the universe. Unlike conventional reheating scenarios that involve specific couplings between the inflation and other fields, gravitational reheating relies on the dynamics of the spacetime itself, making it a universal process that does not depend on the details of particle interactions.

 In fact, at the Lagrangian level, the quantum field is coupled with the Ricci scalar, and since in quintessential inflation the adiabatic evolution is disrupted when the universe transitions from the inflationary epoch to the kination era, massive particles are produced during this phase transition.

The mode–frequency corresponding to heavy, massive particles conformally coupled with gravity is 
\begin{eqnarray}\label{omega}
\omega_k^2(\eta)=k^2+a^2(\eta)m_{\chi}^2,\end{eqnarray}
where $m_{\chi}$ represents the mass of the produced particles. Defining “END” as the end of inflation, occurring when {$\dot{H}=-H^2$} or equivalently when the principal slow-roll parameter $\epsilon$ equals one, if we expand the scale factor up to second order around this point using Taylor expansion, we obtain
\begin{eqnarray}\label{aquadrat}
a^2(\eta)\cong a_{\rm END}^2+2a_{\rm END}^3H_{\rm END} (\eta-\eta_{\rm END})
\nonumber\\ 
+{2}a_{\rm END}^4H_{\rm END}^2 (\eta-\eta_{\rm END})^2 .
\end{eqnarray}
Then, inserting this last  expression in \eqref{omega}, the frequency $\omega_k(\eta)$ can be approximated, up to order two around $\eta_{\rm END}$,  by
\begin{eqnarray}
\omega_k^2(\eta)
\cong  k^2 +\frac{m_{\chi}^2a_{\rm END}^2}{2}
+2 a_{\rm END}^4H_{\rm END}^2m_{\chi}^2
 \nonumber\\ 
\times 
\Bigg(\eta-\eta_{\rm END}+\frac{1}{2a_{\rm END}H_{\rm END}} \Bigg)^2. 
\end{eqnarray}
For this frequency, the $\beta$-Bogoliubov coefficient is
\begin{eqnarray}
       |\beta_k|^2= \exp\Big(-\tfrac{\pi\big(k^2+
       \tfrac{a^2_{\rm END}m^2_{\chi}}{2} \big)}{\sqrt{2}a^2_{\rm END}m_{\chi}H_{\rm END}}\Big),
   \end{eqnarray}
and the corresponding energy density at the onset of kination 
\cite{deHaro:2022ukj}
is given by
\begin{eqnarray}\label{rho1}
   \langle \rho_{kin}\rangle
     \cong
   \frac{1}{4\pi^3}
   e^{-\frac{\pi m_{\chi}}{2\sqrt{2}H_{\rm END}}}
   \sqrt{\frac{m_{\chi}}{\sqrt{2}H_{\rm END}}} H_{\rm END}^2m_{\chi}^2, \end{eqnarray}
demonstrating that the energy density of the produced particles decreases exponentially for masses greater than $H_{\rm END}$.
 To achieve a reheated universe,
 these particles have to decay in lighter ones.
Then, 
using the {Stefan–Boltzmann} law $\rho_{\rm reh}=\frac{\pi^2 g_{\rm reh}}{30}T_{\rm reh}^4$, where {$g_{\rm reh}\cong 107$} are the effective degrees of freedom for the {standard model}, the reheating temperature is given by {(see \cite{AresteSalo:2017lkv,deHaro:2022ukj} for details): }
\begin{eqnarray}\label{reheating1}
 T_{\rm reh}= 
  \left(\frac{10}{3\pi^2g_{\rm reh}} \right)^{1/4}
 \left(\frac{\langle\rho_{kin}\rangle^3}{H_{\rm END}^3
 \Gamma M_{pl}^8}\right)^{1/4}M_{pl},
 \end{eqnarray} 
where the decay rate,   { $\Gamma\sim \frac{h^2m_{\chi}}{8\pi}$},   has to satisfy
\begin{eqnarray}\label{bound}
\frac{\langle\rho_{kin}\rangle}{3H_{\rm END}M_{pl}^2}\leq \Gamma\leq  H_{\rm END},
\end{eqnarray}
which results from the fact that $\Gamma \leq H_{\rm kin} \approx H_{\rm END}$ (indicating that decay occurs after the onset of kination) and $\langle \rho_{\rm dec} \rangle \leq 3 \Gamma^2 M_{\rm pl}^2$ (indicating that decay occurs before the end of kination). The maximum reheating temperature, denoted as $T_{\rm reh}^{\rm max}$, is achieved when decay occurs at the end of kination. This is because, throughout the kination phase, the energy density of the produced particles scales as $a^{-3}$ (and after decay, it scales as $a^{-4}$). Consequently, it quickly matches the energy density of the inflation. Therefore, by choosing $\Gamma = \frac{\langle \rho_{\rm kin} \rangle}{3 H_{\rm END} M_{\rm pl}^2}$ and using the energy density of the produced particles (\ref{rho1}), we can determine the maximum reheating temperature:
 \begin{eqnarray}\label{tmax0}
 T_{{\rm reh}}^{{\rm max}}(m_{\chi})
   \cong \frac{1}{5\pi^2} e^{-\frac{\pi m_{\chi}}{4\sqrt{2}H_{\rm END}}}
  \sqrt{\frac{H_{\rm END}}{M_{\rm pl}}} m_{\chi}.  
 \end{eqnarray}  
 {The maximum reheating temperature depends on the mass of the particles, reaching its peak value when $m_{\chi} = \frac{4 \sqrt{2}}{\pi} H_{\rm END}$. At this mass, the reheating temperature is approximately}
\begin{eqnarray}
 T_{{reh}}^{{max}}\left(\frac{4\sqrt{2}}{\pi}H_{END}\right)\cong  
 \frac{4\sqrt{2}}{5\pi^3e}\sqrt{\frac{H_{END}}{M_{pl}}}H_{END}.
\end{eqnarray}
Then, for a typical Hubble rate at the end of inflation, approximately $H_{END} \sim 10^{-6} M_{pl}$, the maximum value is
   \begin{eqnarray}
T_{{\rm reh}}^{{\rm max}}\left(2\times 10^{-6} M_{\rm pl}\right)
 \cong 2\times 10^7\mbox{ GeV}.
\end{eqnarray}

\subsection{Application to $\alpha$-Attractors}

{In the context of $\alpha$-attractors (refer to the potential in (\ref{alpha})), we determine $H_{END}$ analytically by computing the slow-roll parameter:}

\begin{eqnarray}
    \epsilon=\frac{M_{pl}^2}{2}\left( \frac{V_{\varphi}}{V} \right)^2=\frac{n^2}
    {12\alpha}\frac{1}{\cosh^4\left(\frac{\varphi}{\sqrt{6\alpha}M_{pl}} \right)}.
\end{eqnarray}

Since inflation ends when $\epsilon=1$
and
noting  that
$
\mathop{\mathrm{arccosh}}(x)=\ln(x-\sqrt{x^2-1}),
$
one has
\begin{eqnarray}
    \varphi_{END}=\sqrt{6\alpha}\ln\left( \frac{\sqrt{n}}{(12\alpha)^{1/4}}-\sqrt{\frac{n}{\sqrt{12\alpha}}-1}
    \right) M_{pl},
\end{eqnarray}
and thus, 
\begin{eqnarray}
V(\varphi_{END})
\cong 
\alpha 
10^{-10} M_{pl},
\end{eqnarray}
where, after using  that $\rho_{END}=\frac{3V(\varphi_{END})}{2}$, one obtains
\begin{eqnarray}
    H_{END}\cong 
    \sqrt{\frac{\alpha}{2}} 10^{-5} M_{pl}.\end{eqnarray}

\begin{figure}
   \includegraphics[width=0.4\textwidth]{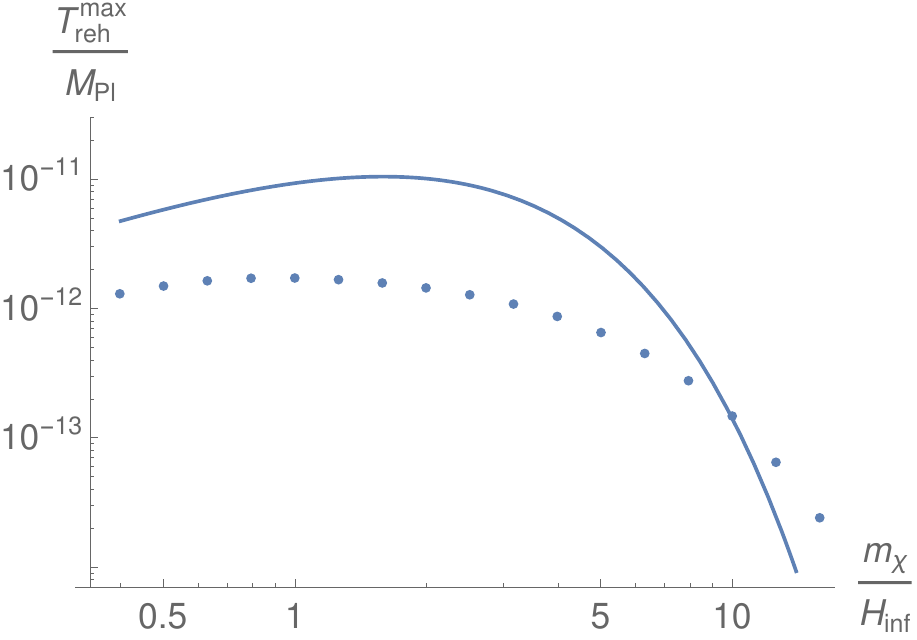}
   \caption{Numerical (in dots) and analytic
   values for  the maximum reheating temperature.
   The values of the parameters are $H_{inf}=10^{-6} M_{pl}$, $\alpha=10^{-2}$, and $n=124$. The figure has been taken from \cite{deHaro:2022ukj}.    }
   \label{fig:beta}
\end{figure}

Therefore,  the maximum reheating temperature in the conformally coupled case   is given by 
\begin{align}
T_{{reh}}^{{max}}(m_{\chi})\cong 6\times 10^{-4} 
\left(\frac{\alpha}{2}\right)^{1/8}\exp\left({-\frac{ \pi \times 10^5  m_{\chi}}{4\sqrt{\alpha}M_{pl}}}\right)
 \nonumber\\ \times 
 \left(\frac{m_{\chi}}{M_{pl}} \right)^{1/4} m_{\chi},
\end{align}
which, for ${\alpha=10^{-2}}$, becomes
   \begin{align}\label{maximumtemperature}
T_{{reh}}^{{max}}(m_{\chi})\cong 3\times 10^{-4} 
\exp\left(-\frac{ \pi\times 10^6  m_{\chi}}{4M_{pl}} \right)
\left(\frac{m_{\chi}}{M_{pl}} \right)^{1/4}
m_{\chi},
\end{align}
which, as we can see in Figure \ref{fig:beta}, matches very well with the numerical results.

\subsection{Relation Between the Number of Last  $e$-Folds and the Reheating Temperature: $\alpha$-Attractors}

\begin{figure}
\includegraphics[width=0.4\textwidth]{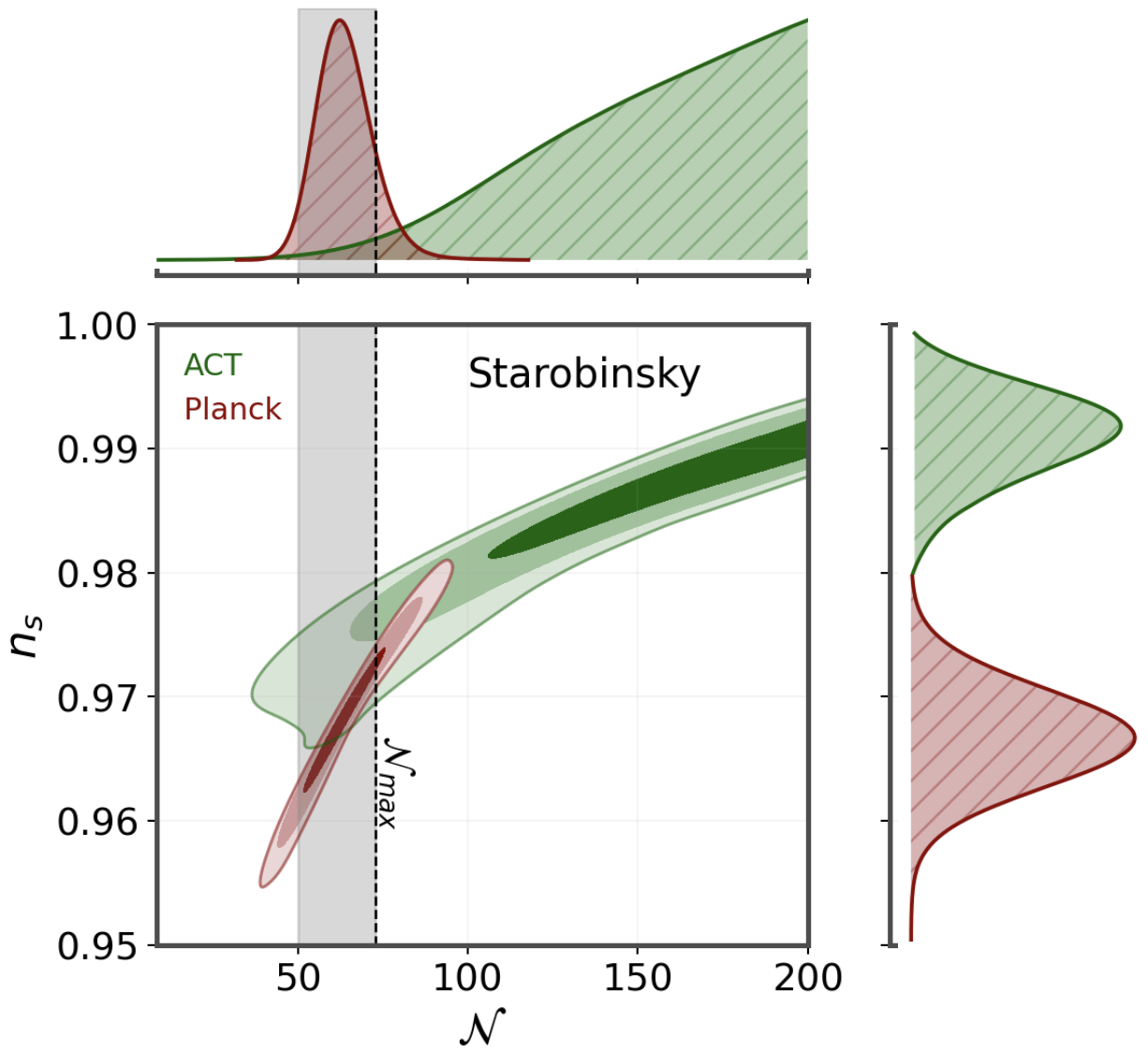}
   \caption{\small {Two-dimensional contours at 68\%, 95\%, and 99\% confidence levels  in the ($n_s$, ${N}$) plane for the Starobinsky model. The gray vertical band indicates the typical range of $e$-folds expansion, ${N} \in [50, {N}{max}]$, expected during inflation. The upper bound, ${N}{max} \le 73$, is shown by the black dashed line. This figure is adapted from \cite{Giare:2023wzl}.}
  }
    \label{fig:starobinsky}
\end{figure}

The well-known formula that relates the number of $e$-folds with the reheating temperature and the main slow-roll parameter is as follows
(see \cite{deHaro:2023xcc} for details):
\begin{align}\label{Nxx}
    N(T_{\rm reh}, \epsilon_*)
    \cong 54.47 +\frac{1}{2}    \ln \epsilon_*  +\frac{1}{3}\ln \left( \frac{M_{\rm pl}^2}{T_{\rm reh}H_{\rm END}}\right),   \end{align} 
    where the $\ast$ means that the quantities are evaluated at the {horizon crossing}.
Note that, to obtain this formula, we have
{disregarded}
 the term $\ln\left(\frac{a_{\rm END}}{a_{\rm kin}} \right)$ since 
 it is close to zero, and we have
 {used that, in practice,  the energy density  does not change}
  during the phase transition from the end of inflation to the {onset} of kination.
On the other hand, 
for $\alpha$-attractors, one has 
\begin{eqnarray}\label{spectral}
    N(n_s)\cong \frac{2}{1-n_s} \quad \mbox{and} \quad \epsilon_*(n_s)\cong \frac{3\alpha}{16}(1-n_s)^2.
\end{eqnarray}
Equaling (\ref{Nxx}) and (\ref{spectral}),
 one finds the reheating temperature as a function of the spectral index as follows:
\begin{align}\label{reheating-alpha}
    T_{reh}\cong \alpha (1-n_s)^2 \exp\left(169
    +\frac{\sqrt{3\alpha}}{2}    -\frac{6}{1-n_s}\right) M_{pl}.
\end{align}
Choosing, for example,  ${\alpha=10^{-2}}$, the reheating temperature will become
\begin{align} \label{alpha-0.01}    
T_{\rm reh}\cong  (1-n_s)^2 \exp\left(169+\frac{\sqrt{3}}{20}
      -\frac{6}{1-n_s}\right) 10^{-2} M_{\rm pl},
\end{align}
and  the allowed values of {$n_s$}, which {ensure} a reheating temperature {preserving} with the {BBN success}
($T_{\rm reh}\geq 1$ MeV) and  {overpassing}  the gravitino problem  ($T_{\rm reh}\leq 10^9$ GeV), belong in the range  {$(0.9665,0.9709)$},
which enters in the {$95\%$} confidence level of the observational data provided by Planck's team \cite{Planck:2018vyg}, because 
 {$n_s=0.9649\pm 0.0042$}.
In addition, the { tensor/scalar ratio}
satisfies $r=16\epsilon_*=3\alpha(1-n_s)^2$, and thus, it is constrained  by  {$0.000024<r<0.000034$}.
In particular, considering instant preheating, we have seen that the reheating temperature belongs to the domain
$10^6 \mbox{ GeV}\leq T_{reh}\leq 10^9 \mbox{ GeV}$, which constrains the spectral index residing in the narrow interval
$(0.9667,0.9677)$.
However, the 
{Atacama Cosmology Telescope} (ACT) \cite{ACT:2020frw}
  provides {$n_s=0.9858_{-0.0030}^{+0.0051}$}, 
  which  disfavors the $\alpha$-attractors. 
  We can see in   Figure \ref{fig:starobinsky}
  that the same happens for the other inflationary potentials, such as Starobinsky and SUSY inflationary models 
  (details can be found in \cite{Giare:2023wzl}).

\subsubsection*{Relation Between 
{ $n_s$ and $m_{\chi}$ }}

{Using the earlier result for the maximum reheating temperature (\ref{maximumtemperature}) with {$\alpha=10^{-2}$}, which solely depends on the mass of the produced particles, we derive the relationship between the spectral index and the mass of the particles produced:}
\begin{align}\label{alpha-alpha}
    \frac{2.6057}{1-n_s}-1.75\log(1-n_s)=79.301+\frac{1.2398}{1-n_s}X-1.25\log X,
\end{align}
where we have introduced the notation $X\equiv 10^4 \frac{m_{\chi}}{M_{pl}}$. 
{For the minimum value of the spectral index, Equation  (\ref{alpha-alpha}) only has one solution,
$X_{min}=\frac{1.25(1-n_s)}{1.2398}$},  
and 
by inserting it
 into (\ref{alpha-alpha}), we obtain
\begin{eqnarray}\label{xxxx}
    \frac{2.6057}{1-n_s}-0.5\log(1-n_s)=80.5465.
    \end{eqnarray}
The only solution to this equation is {$\bar{n}_s\cong0.9673$}
because the function
\begin{eqnarray}
\frac{2.6057}{1-n_s}-0.5\log(1-n_s)
\end{eqnarray}
is increasing. 
{ And thus}, for this minimum value of the spectral index, Equation (\ref{alpha-alpha}) also has a unique solution: {$m_{\chi}\cong 10^{-6}M_{pl}$}, which leads to a maximum reheating temperature  around {$10^7$ GeV}.  For values of {$n_s$} in the {domain}  $(0.9673, 0.9709)$ where {$n_s=0.9709$} is the maximum allowed value (recall the Planck 2018 data $n_s=0.9649\pm 0.0042$ \cite{Planck:2018jri}),  leading to a reheating temperature above {$1$ MeV}, 
Equation (\ref{alpha-alpha}) always has two solutions. For example, when $n_s=0.9709$, 
{ we find}
{$m_{\chi}\cong 3\times 10^{-5} M_{\rm pl}$} and {$m_{\chi}\cong 5\times 10^{-15}M_{\rm pl}$}, with a reheating temperature 
{around}
 $1$ MeV. In other words,
 {
 if the decay happens at the end of kination (resulting in the maximum reheating temperature), the spectral index must fall within the range $(0.9673, 0.9709)$, with two possible masses corresponding to each value. Alternatively, for masses within the range $5\times 10^{-15}\leq m_{\chi}/M_{\rm pl}\leq 3\times 10^{-5}$, there is a corresponding spectral index between $0.9673$ and $0.9709$, leading to a viable maximum reheating temperature. 
  }

\section{$H_0$ Tension}
\label{sec-H0}

{The $\sim 5\sigma$ discrepancy between Planck (within $\Lambda$CDM) \cite{Planck:2018vyg} and SH0ES \cite{Riess:2021jrx} is a serious issue at the present moment, which suggests a possible revision of the $\Lambda$CDM cosmology. Various mechanisms are available to address the $H_0$ discrepancy \mbox{(see ~\cite{DiValentino:2021izs,Perivolaropoulos:2021jda,Abdalla:2022yfr}).} Such modifications are either performed at the late time or during the early evolution of the universe; however, the actual truth is still unknown ~\cite{DiValentino:2021izs,Perivolaropoulos:2021jda,Abdalla:2022yfr}.  }
Adhering to the insights provided in \cite{Poulin:2023lkg} (refer to the beginning of Section II of \cite{Poulin:2023lkg}), it is elucidated that the parameter instrumental in constraining $H_0$ through observations of early universe physics is the angular scale of the sound horizon:
\begin{eqnarray}
\theta_s(z_*)=\frac{r_s(z_*)}{D_A(z_*)},
\end{eqnarray}
where $z_*$ is the redshift at the baryon–photon decoupling and $D_A(z)$,  
the  angular diameter distance, is defined by
\begin{eqnarray}
D_A(z)=\int_0^z\frac{dz'}{H(z')},
\end{eqnarray}
and 
the  physical size of the sound horizon at the redshift $z_*$
is
\begin{equation}
r_s(z_*) = \int_{z_*}^{\infty} \frac{c_s(z)}{H(z)} dz.
\end{equation}
The speed of sound in the baryon–photon fluid is given by $c_s(z) \approx \left(3 + \frac{9\rho_b(z)}{4\rho_{\gamma}(z)}\right)^{-1/2}$, where $\rho_b(z)$ and $\rho_\gamma(z)$ represent the densities of baryons and photons at redshift $z$, respectively.
Note that we can write
\begin{eqnarray}
    c_s(z)= \left( 3\left(1+\frac{3\Omega_{\rm b,0}h^2}{4\Omega_{\rm \gamma,0}h^2}\frac{1}{1+z} \right)\right)^{-1/2},
\end{eqnarray}
and using the data $\Omega_{\rm b,0}h^2\cong 0.02237$ and 
$\frac{3}{4\Omega_{\rm \gamma,0}h^2}\cong 31500$
\cite{Chen:2018dbv}, 
for $z\geq z_*$, one can safely make the approximation $c_s(z)\cong c_s(z_*)\cong 0.45$. In addition, for the $\Lambda$CDM model, denoting by $H_{\Lambda}$ the Hubble rate for this model,   one has 
\begin{align}
D_A^{\Lambda}(z_*)=\frac{\sqrt{3}}{M_{pl}}
\int_0^{z_*}\frac{dz}{\sqrt{{\rho}_{\Lambda}+
{\rho}_{\rm m,0}(1+z)^3\left( 1+\frac{1+z}{1+z_{\rm eq}}\right)}}
\end{align}
which, at the time of baryon–photon decoupling, turns out to be $\cong  5.277\times 10^{60} M_{\rm pl}^{-1}$
under the assumption of 
 \begin{eqnarray}
\rho_{m,0}\cong 3.2877\times 10^{-121} M_{pl}^4, \quad
z_*=1089.8,  \quad z_{\rm eq}=3387, \nonumber\\ 
H_{\Lambda,0}=67.66~\mbox{km/s/Mpc}\cong 5.9356\times 10^{-61} M_{\rm pl}.
\end{eqnarray}
Note that here, $\rho{\rm m, 0}$ denotes the present-day value of the matter density. 
Recall that, in the $\Lambda$CDM model, since the dark energy is denoted by $\rho_{\Lambda}=\Lambda M_{\rm pl}^2$, and since 
\begin{eqnarray}
3H_{\Lambda,0}^2M_{\rm pl}^2=\rho_{\Lambda}+\frac{2+z_{\rm eq}}{1+z_{\rm eq}}\rho_{\rm m,0},
\end{eqnarray}
we obtain $\rho_{\Lambda}
\cong 7.3053\times 10^{-121}M_{pl}^4.$

{
\subsection{$H_0$ Tension in Quintessential Inflation: Inclusion of a Dynamical DE}}

Dealing with the exponential 
$\alpha$-attractor
\begin{eqnarray}\label{alpha1}
V(\varphi)=\lambda M_{\rm pl}^4e^{-n\tanh\left(\frac{\varphi}{\sqrt{6\alpha}M_{\rm pl}} \right)}, \qquad  n\equiv \kappa\sqrt{6\alpha},
\end{eqnarray} 
{one approach to resolving the $H_0$ tension in quintessential inflation is to modify the expansion history for {$z \lesssim 2$}, adjusting the value of $H_0$ without altering the {angular diameter distance} at the time of {last scattering}. Since both quintessential inflation (for an exponential $\alpha$-attractor) and {$\Lambda$CDM} predict the same value of $r_s(z_*)$  $-$ as prior to the {recombination epoch} at {$z_*$}, the Hubble rate is mainly governed by matter and radiation energy densities—both models should yield the same {angular diameter distance}. However, since the energy density of the scalar field in quintessential inflation decreases, this match cannot be achieved. Therefore, to resolve the problem, one possible solution is to introduce a dynamical DE  to modify the evolution of the universe at {$z \lesssim 2$}. The dynamical DE is not the only possibility to alleviate the $H_0$ tension; there are various ways to modify the expansion history of the universe \cite{DiValentino:2021izs}. However, at this moment, there is no perfect cosmological model that can solve the $H_0$ tension in agreement with other astronomical surveys. The dynamical DE is a very natural extension to the cosmological constant.  }

To follow this direction, 
we can consider a very well known {dynamical} DE EoS parameter \cite{Chevallier:2000qy,Linder:2002et}:
\begin{eqnarray}
    w_{\rm de}(z)=w_0+\frac{z}{1+z} w_a.
\end{eqnarray}
where $w_0$ is the present value of $w_{\rm de}$, and $w_a$, a constant, characterizes the dynamical nature of $w_{\rm de}$. For example, ${w}_a =0$ represents that $w_{\rm de}$ does not evolve with time. 
Then, using the conservation equation of the DE fluid
\begin{eqnarray}
    \frac{d\rho}{dz}=\frac{3}{1+z}\left(1+w_0+{w}_a-\frac{w_a}{1+z} \right)\rho,
\end{eqnarray}
one can find the evolution of the energy density as 
\begin{eqnarray}
    \rho_{\rm de}(z)=\rho_{\rm de,0}(1+z)^{3(1+w_0+w_a)}e^{-3w_a\frac{z}{1+z}}.
\end{eqnarray}
Therefore, we obtain
\begin{eqnarray}
D_A^{\varphi+\rm de}(z_*)= \frac{\sqrt{3}}{M_{\rm pl}}
\int_0^{z_*} dz \Bigg[{\rho}_{\varphi,0}+
{\rho}_{\rm de}(z) 
\nonumber\\
+
{\rho}_{\rm m,0}(1+z)^3\left( 1+\frac{1+z}{1+z_{\rm eq}}\right) \Bigg]^{-\frac{1}{2}},
\end{eqnarray}
where we now have three free parameters, namely $\rho_{\rm de,0}$, $w_0$, and $w_a$, with the constraint 
${\rho}_{\varphi,0}+{\rho}_{\rm de,0}>
{\rho}_{\Lambda}$ in order to increase the present value of the Hubble rate.
Choosing {${\rho}_{\varphi,0}+{\rho}_{de,0}=9.1\times 10^{-121} M_{pl}^4$}, where $\rho_{\varphi}$ denotes the energy density of the $\varphi$-field,  this leads to {$H_{\varphi+de,0}=6.4259\times 10^{-61} M_{pl}\cong 73.25 \mbox{km/s/Mpc}$}. 

{ 
Imposing 
$D_A^{\varphi+\rm de}(z_*)\cong D_A^{\Lambda}(z_*)
\cong {5.277\times 10^{60} M_{\rm pl}^{-1}},$
we have obtained the following values:
\begin{enumerate}
    \item Considering ${\rho}_{\varphi,0}=7\times 10^{-121}M_{\rm pl}^4$, one can approximate $w_0\cong -4.6333$ and $w_a=-0.0333$. The approximate value of $w_0$ corresponds to a super phantom stage of the dark energy at the present epoch. However, we note that such a high negative value of $w_0$ arises because of the choice of $\rho_{\varphi, 0}$.  
On the other hand, the present value of the effective EoS parameter $w_{0}^{\rm eff}$, which is encoded in the quintessential inflation field and the phantom field, is    
\begin{eqnarray}
    w_{0}^{\rm eff}=\frac{-{\rho}_{\varphi,0}+w_0 {\rho}_{\rm de,0}}{\rho_{\varphi,0}+\rho_{\rm de, 0}}\cong 
     -1.8384 .
\end{eqnarray}    
In addition,  ${\rho}_{\rm de}(z_*)\cong 10^{-154}M_{\rm pl}^4$, ${\rho}_{\rm de}(2)\cong 2\times 10^{-127}M_{\rm pl}^4$, {${\rho}_{\rm de} \rightarrow 0$ as $z \rightarrow -1$.}

\item  When considering ${\rho}_{\varphi,0}=6\times 10^{-121}M_{\rm pl}^4$, one can approximate $w_0\cong -1.9333$ and ${w}_a=-0.0333$.
In that case,  $ w_{0}^{\rm eff}=-1.3179$,  ${\rho}_{\rm de}(z_*)\cong 3\times 10^{-130}M_{\rm pl}^4$, $\bar{\rho}_{\rm de}(2)\cong 10^{-122}M_{\rm pl}^4$, {${\rho}_{\rm de} \rightarrow 0$ as $z \rightarrow -1$.}

\item On the other hand, if ${\rho}_{\varphi,0}=5\times
10^{-121}M_{\rm pl}^4$, $w_0\cong -1.5333$ and ${w}_a=-0.0333$.
In that case, $ w_{0}^{\rm eff}=  -1.2402$,  ${\rho}_{\rm de}(z_*)\cong 3\times 10^{-128}M_{\rm pl}^4$, ${\rho}_{\rm de}(2)\cong 10^{-122}M_{\rm pl}^4$, {${\rho}_{\rm de} \rightarrow 0$ as $z \rightarrow -1$.}
    \end{enumerate}
}

{Thus, it is clear to understand that depending on the strength of ${\rho}_{\varphi,0}$, the nature of $w_0$ and the effective EoS at the present epoch alter. This is a clear indication that depending on ${\rho}_{\varphi,0}$, one can expect a quintessential nature of $w_0$; however, the phantom behavior of DE can increase the expansion rate of the universe, and as a result of which, one can mitigate the $H_0$ tension~\cite{DiValentino:2021izs}. }
Finally, since {${w}_a<0$},  for {$z<0$}, hence, the effective EoS parameter of the total fluid, namely, $w^{\rm eff}(z)$, will eventually become positive, that is, the Big Rip singularity \cite{deHaro:2023lbq}, a typical nature 
of phantom fluids, is forbidden because the fluid becomes non-phantom.  {To conclude this section, we would like to emphasize that addressing the $H_0$ tension may require going beyond general relativity. In particular, the inclusion of a dynamical DE in the context of $f(R)$ gravity~\cite{Montani:2023xpd} or the introduction of the matter creation process in modified gravity \cite{Montani:2024xys} can play an effective role in reconciling the $H_0$ tension.  }

\subsection{Quintessential Inflation as a Source of EDE}

{
Another popular alternative to reconcile the 
Hubble tension involves EDE \cite{Poulin:2018cxd,Sakstein:2019fmf,Niedermann:2020dwg,Ivanov:2020ril,Kamionkowski:2022pkx,Poulin:2023lkg,Efstathiou:2023fbn,Seto:2024cgo,Giare:2024akf}.  
The presence of EDE can reduce the size of the sound horizon at recombination by adding extra energy density before recombination. 
This increase in pre-recombination energy boosts the Hubble rate at the time of photon decoupling. 
As a result, to preserve consistency with the concordance model—where the angular scale stays unchanged—the present-day Hubble rate must be higher than in the $\Lambda$CDM model\footnote{However, at this point, the readers might be interested to know that all early-time modifications may not be sufficient to solve the Hubble tension}~\cite{Jedamzik:2020zmd,Vagnozzi:2023nrq})}. 

\begin{figure}
\includegraphics[width=0.4\textwidth]{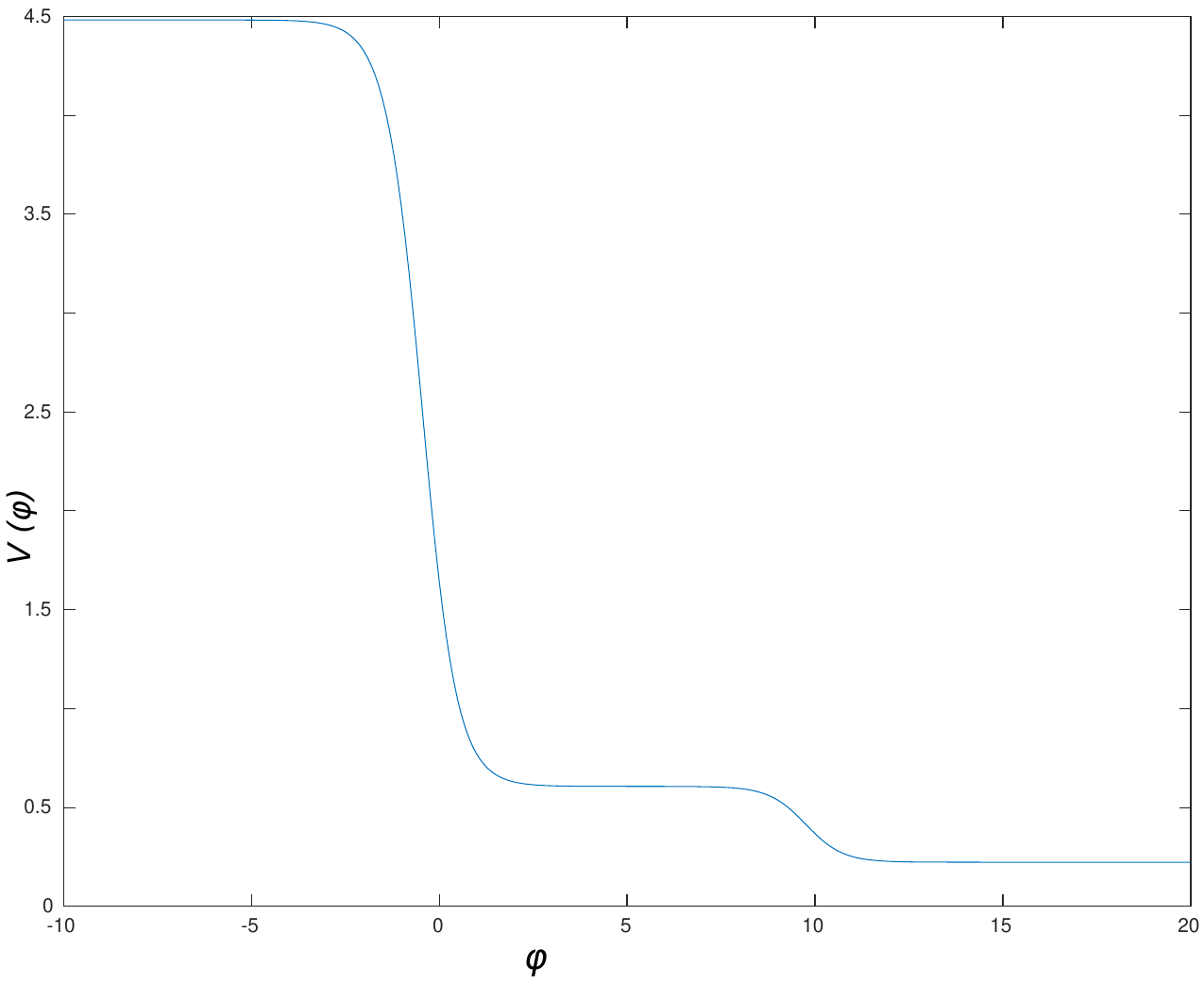}
	\caption{Plot of  the potential (\ref{alpha_improved}) as a function of the field, taking some typical values of the parameters involved, namely, $\alpha=1/6$,  $\bar{\lambda}=1$,  $n=1$, $m=1/2$, and $\varphi_c=10 M_{pl}$. }\label{fig2}
\end{figure}

In quintessential inflation, to inject this extra energy density,
we improve the exponential potential that appears in the Lagrangian (\ref{lagrangian}) as follows:
\begin{eqnarray}
    V(\phi)=\bar{\lambda} e^{-\kappa\phi}e^{-\bar{\kappa}\frac{\phi-\phi_c}{6\alpha-\phi\phi_c}}
    M_{pl}^4,
\end{eqnarray}
which,
in terms of the canonical field $\varphi$ defined in (\ref{canonical}), leads to the following (see Figure \ref{fig2}): \begin{eqnarray}\label{alpha_improved}
V(\varphi)=\bar{\lambda} e^{-n\tanh\left(\frac{\varphi}{\sqrt{6\alpha}M_{pl}} \right)}
e^{-m\tanh\left(\frac{(\varphi-\varphi_c)}{\sqrt{6\alpha}M_{pl}} \right)}
M_{pl}^4,
\end{eqnarray}
where  $m=\frac{\bar{\kappa}}{\sqrt{6\alpha}}$ and $\phi_c=\sqrt{6\alpha}\tanh\left(\frac{\varphi_c}{\sqrt{6\alpha} M_{pl}} \right)$, with
$\varphi_c
\cong \left(-10+2\sqrt{\frac{2}{3}}\ln\left( \frac{M_{pl}}{T_{\rm reh}}\right)\right)M_{pl}$,
is the value of the scalar field close to the matter–radiation equality, 
as has been  shown in 
\cite{deHaro:2021swo}.
In fact, the phase transition at $\varphi_c$ has to occur close to the recombination.
Note that, 
at early times, the potential behaves like (\ref{alpha1}) with $\lambda=\bar{\lambda}e^m$.
Well before the matter–radiation equality, 
the field freezes during the radiation phase, and the potential becomes 
$V(\varphi)=\bar{\lambda} e^{-(n-m)}M_{pl}^4$ with $w_{\varphi}\cong -1$, i.e., it acts as a cosmological constant; this becomes an injection of energy as in the EDE models, and soon after the field is greater than $\varphi_c$, it rolls down the potential, and  the EoS parameter of the field goes to $1$ because the field enters in a second  kination phase, which disappears at the present time because the field freezes once again, recovering $w_{\varphi}=-1$, and the potential has the  
form $V(\varphi)=\bar{\lambda} e^{-(n+m)} M_{pl}^4$, becoming another smaller  cosmological constant.
More precisely, the model, after reheating,  has to be understood as a dynamical cosmological constant, which changes its scale close to the recombination.

\subsubsection{Qualitative Calculations}

First of all, dealing with the concordance model,  we write
\begin{eqnarray}
    H(z\leq z_*)\cong H_{\Lambda, 0}\sqrt{\Omega_{\Lambda}+\Omega_{\rm m,0}^{\Lambda}(1+z)^3},
\end{eqnarray}
 where we have introduced the notation $\Omega_{\rm m,0}^{\Lambda}=\frac{\rho_{\rm m,0}}{3H_{\Lambda, 0}^2 M_{\rm pl}^2}$,  $\Omega_{\Lambda}=\frac{\rho_{\Lambda}}{3H_{\Lambda, 0}^2 M_{\rm pl}^2}$, and we have disregarded the radiation term, which is negligible for $z\leq z_*$.  Thus, 
\begin{eqnarray}
    D_A^{\Lambda}(z\leq z_*)\cong 
    \frac{C}{H_{\Lambda, 0} \Omega_{\Lambda}^{1/6}
    (\Omega_{m,0}^{\Lambda})^{1/3}},
\end{eqnarray}
where 
\begin{eqnarray}
    C=\int_{b_{\Lambda}}^{b_{\Lambda}(1+z_*)}
    \frac{dx}{\sqrt{1+x^3}}\quad
    ,\nonumber\\
    \mbox{with}\quad 
    b_{\Lambda}\equiv \left(\frac{\Omega_{\rm m,0}^{\Lambda}}{\Omega_{\Lambda}} \right)^{1/3}\cong 0.75.
\end{eqnarray}
On the other hand, for $z_{\rm eq}\gg z\geq z_*$, one has
$\Omega_{\rm m,0}^{\Lambda}(1+z)^3\gg \Omega_{\Lambda}$, and thus
\begin{eqnarray}
    H(z)\cong H_{\Lambda, *}\sqrt{\Omega_{m,*}^{\Lambda}}\left( \frac{z}{z_*}\right)^{3/2},
\end{eqnarray}
where $H_{\Lambda,*}=H_{\Lambda}(z_*)$ and $\Omega_{\rm m,*}^{\Lambda}=\frac{\rho_{\rm m,0}(1+z_*)^3}{3 H_{\Lambda,*}^2 M_{pl}^2}$. 
Therefore, since,  
for $z\geq z_*$, one has $c_s(z)\cong c_s(z_*)$, one can make the approximation
\begin{eqnarray}
    r_s(z_*)\cong \frac{2c_s(z_*) z_*}{H_{\Lambda,*}\sqrt{\Omega_{m,*}^{\Lambda}}},
\end{eqnarray}
and using the relationship $\Omega_{m,*}^{\Lambda}\cong \Omega_{m,0}^{\Lambda} \left(\frac{H_{\Lambda,0}}{H_{\Lambda, *}} \right)^2 z_*^3$, 
we obtain the following expression of the angular scale of the sound horizon:
\begin{eqnarray}\label{angle_Lambda}
    \theta( z_*)\cong \frac{2c_s(z_*)}{C\sqrt{z_*}}\left(\frac{\Omega_{\Lambda}}{\Omega_{m,0}^{\Lambda}} \right)^{1/6}.
\end{eqnarray}
Dealing with quintessential inflation, in the same way as for the concordance model, we have 
\begin{eqnarray}
    D_A^{\varphi}(z\leq z_*)\cong 
    \frac{\bar{C}}{H_{\varphi,0} \Omega_{\varphi,0}^{1/6}
    (\Omega_{m,0}^{\varphi})^{1/3}},
\end{eqnarray}
where
\begin{eqnarray}
    \bar{C}=\int_{b_{\varphi}}^{b_{\varphi}(1+z_*)}
    \frac{dx}{\sqrt{1+x^3}}\quad \mbox{with}\quad 
    b_{\varphi}=\left(\frac{\Omega_{m,0}^{\varphi}}{\Omega_{\varphi,0}} \right)^{1/3},
\end{eqnarray}
 and we have introduced the notation $\Omega_{\rm m,0}^{\varphi}=\frac{\rho_{\rm m,0}}{3H_{\varphi, 0}^2 M_{\rm pl}^2}$,
 $\Omega_{\varphi,0}=\frac{\rho_{\varphi,0}}{3H_{\varphi, 0}^2 M_{\rm pl}^2}$. 
On the other hand, for $z_*\leq z\ll z_{\rm eq}$, we have 
\begin{eqnarray}
    H(z)\cong H_{\varphi,*}\sqrt{\Omega_{\varphi,*}+\Omega_{\rm m,*}^{\varphi}
    \left(\frac{1+z}{1+z_*}\right)^3},
\end{eqnarray}
and thus,
\begin{eqnarray}
    r_s(z_*)\cong \frac{c_s(z_*) z_* D}{H_{\varphi,*} \Omega_{\varphi,*}^{1/6}
    (\Omega_{\rm m,*}^{\varphi})^{1/3}},
\end{eqnarray}
where
\begin{eqnarray}
    D=\int_{b_{\varphi}}^{\infty}\frac{dx}{\sqrt{1+x^3}}\qquad \mbox{with}\qquad
    b_{\varphi}=\left( \frac{\Omega_{\rm m,*}^{\varphi}}{\Omega_{\varphi,*}}\right)^{1/3}.
\end{eqnarray}
Now, using 
$\Omega_{\rm m,*}^{\varphi}\cong \Omega_{\rm m,0}^{\varphi} \left(\frac{H_{\varphi,0}}{H_{\varphi,*}} \right)^2 z_*^3,
$ we obtain 
\begin{eqnarray}\label{angle_quintessential inflation}
    \theta( z_*)\cong \frac{Dc_s(z_*)}{\bar{C}}\left(\frac{\rho_{\varphi,0}}{\rho_{\varphi,*}} \right)^{1/6}.
\end{eqnarray}
Next, equaling the corresponding angular scales, i.e., 
Equation (\ref{angle_Lambda}) with Equation (\ref{angle_quintessential inflation}),  we obtain
\begin{eqnarray}
    \rho_{\varphi,*}\cong \frac{z_*^3}{64}\left(\frac{DC}{\bar{C}}\right)^6\frac{\Omega_{\rm m,0}^{\Lambda}}{1- \Omega_{m,0}^{\Lambda}}\rho_{\varphi,0},
\end{eqnarray}
where, taking into account that
$C\cong \bar{C}$, $\Omega_{\rm m,0}^{\Lambda}\cong 0.3$, $z_*\cong 1089$ and making the approximation
$D\sim \int_1^{\infty}x^{-3/2}=2$,  we have
\begin{eqnarray}
    \rho_{\varphi,*}\sim 5\times 10^8 \rho_{\varphi,0} \Longleftrightarrow
    {V}_*\sim 5\times 10^8{V}_0.
\end{eqnarray}
Choosing 
$V_0\cong 9.1\times 10^{-121}M_{\rm pl}^4$
to obtain $H_{\varphi,0}\cong 73~\mbox{km/s/Mpc}$, we have
 ${V}_*\sim 5\times 10^{-112}M_{\rm pl}^4.$
That is, 
\begin{eqnarray}
    \bar{\lambda} e^{-(n+m)}\sim 9\times 10^{-121},\qquad \bar{\lambda} e^{-(n-m)}\sim 5\times 10^{-112},\end{eqnarray}
which implies $m\sim 10$.   On the other hand, using the formula of the power spectrum of scalar perturbations \cite{Riotto:2002yw},
we have
\begin{eqnarray}
    \bar{\lambda}\sim 9\alpha (1-n_s)^2\times 10^{-9} e^{-(n+m)},
\end{eqnarray}
and choosing $\alpha\sim 10^{-2}$ and $1-n_s\sim 10^{-2}$, we obtain $n\sim 112$.

\section{Conclusions}
\label{sec-summary}

A unified prescription connecting the early inflationary phase and late quintessential era, in terms of  quintessential inflation,  is the main theme of the present article, where we have focused on two important aspects of modern cosmology, namely, the reheating and the Hubble constant tension. One of the novelties of the present work is to discuss whether the Hubble constant tension can be alleviated within the framework of quintessential inflation, which has received considerable attention in the last couple of years.

We started with the review of the most important reheating mechanisms in quintessential inflation, namely instant preheating and gravitational reheating via the production of heavy particles, obtaining the reheating temperature in terms of the decay rate of these massive particles. After obtaining the reheating temperature, we have related it to the number of $e$-folds, which constrains the spectral index of scalar perturbations. Additionally, in the case of gravitational reheating {and dealing with $\alpha$-attractors coming from super-symmetric gravity theories}, 
we have related the masses of the produced particles with the spectral index, obtaining the viable values of these masses, which are compatible, at the $2\sigma$ confidence level, with the observational values of the spectral index provided by the Planck team \cite{Planck:2018jri}.

Next,  we have analyzed some possible solutions to the $H_0$ tension in the context of quintessential inflation. Dealing with an exponential $\alpha$-attractor representing a quintessential inflation model,  since at late times it acts as a cosmological constant,  we find that the quintessential inflation model alone is not able to increase the $H_0$ value, but
a possibility to alleviate the $H_0$ tension
is  to introduce a dynamical DE  that only acts  at low redshift $z\leq 2$ {with phantom nature at present time}. In addition, for redshifts close to $-1$, the effective nature of the phantom fluid  becomes non-phantom, which prevents the Big Rip singularity that arises in the presence of a phantom fluid~\cite{deHaro:2023lbq}. 
Another possibility is to consider an exponential type of potential (\ref{alpha_improved}) where it acts as a source of EDE. In fact, after reheating, the potential acts as a cosmological constant, which, after a phase transition close to the recombination epoch, decays in another one.

{
In summary, this article emphasizes several key characteristics of models within the framework of quintessential inflation. Notably, addressing the Hubble constant tension is a novel topic in this context. Based on our review, this is the first instance where the question is posed as to whether the quintessential inflation models might provide a potential solution to the Hubble tension. We believe it is crucial to further investigate this issue within this framework using the observational data from various astronomical surveys. }

\begin{acknowledgments}
JdH is supported by the Spanish grants PID2021-123903NB-I00 and RED2022-134784-T
funded by MCIN/AEI/10.13039/501100011033 and by ERDF ``A way of making Europe''. SP acknowledges the financial support from the Department of Science and Technology (DST), Govt. of India under the Scheme   ``Fund for Improvement of S\&T Infrastructure (FIST)'' (File No. SR/FST/MS-I/2019/41).
\end{acknowledgments}

\bibliography{references}

\end{document}